\documentclass{appolb}

\usepackage[title,titletoc]{appendix}
\usepackage[utf8]{inputenc}
\usepackage[square,sort,comma,numbers]{natbib}
\usepackage{graphicx}
\usepackage{lineno}
\usepackage[breaklinks=true]{hyperref}
\usepackage{hypernat}
\usepackage{epsfig}
\usepackage{wrapfig}

\usepackage{graphicx,type1cm,eso-pic,color}

\usepackage{fancyhdr}
\usepackage{pslatex}   
\usepackage{color}
\usepackage{amsmath, amsthm, amssymb} 
\usepackage{endnotes}
\usepackage{xspace}
\usepackage{multirow}

\usepackage{libertine}

\begin{document}
\title{Soft QGP probes with ALICE%
\thanks{Presented at the Cracow Epiphany Conference on the Physics in LHC Run2, Kraków, Poland, January 7-9, 2016.}%
}
\author{Łukasz Kamil Graczykowski
\address{
for the ALICE Collaboration
\\$ $\space $ $
\\Faculty of Physics, Warsaw University of Technology
\\ul. Koszykowa 75, 00-662 Warszawa, Poland
}
}
\maketitle
\begin{abstract}
In heavy-ion collisions at the LHC a hot and dense medium of deconfided partons, the Quark-Gluon Plasma (QGP), is created. Its global properties can be characterized by the measurements of particles in the low transverse momentum (or ``soft") regime, which represent the majority of created particles.
In this report we outline a selection of measurements of the soft probes by the ALICE experiment in pp, p--Pb, and Pb--Pb collisions. The paper focuses on recent flow measurements via angular correlations and femtoscopic studies. The first ever preliminary analysis of $\mathrm{K}^0_{\rm S}\mathrm{K}^{\pm}$ femtoscopy is also presented.

\end{abstract}
\PACS{PACS numbers come here}
  
\section{Introduction}
One of the four large experiments operating at the Large Hadron Collider (LHC)
is A Large Ion Collider Experiment (ALICE)~\cite{Aamodt:2008zz}. It is the experiment dedicated to study the properties and behavior of the strongly interacting matter, the Quark-Gluon Plasma (QGP)~\cite{Shuryak:1978ij}, at the very high temperatures and energy densities reached in ultrarelativistic Pb–Pb collisions.
In addition to heavy ions, ALICE also studies proton-proton and proton-lead collision systems, to provide the baseline for A--A collisions. The LHC Run 1 results from pp and p--A collisions turned out to be as interesting as the results from A--A data, revealing surprising structures previously attributed to the hydrodynamic expansion of the QGP medium. This has triggered the still ongoing debate on potential existence of a collective phase in small systems.

The most unique features of the ALICE detector, allowing measurements of a wide variety of physics phenomena, are the excellent tracking and particle identification (PID) capabilities over a broad momentum range (from just a few MeV/$c$ up to more than 100 GeV/$c$). These capabilities enable studies of both ``soft" (non-perturbative regime of Quantum Chromodynamics, QCD) and ``hard" physics (perturbative regime of QCD).

In this report we focus on selected particle correlation measurements from the soft sector of QCD, describing the bulk properties of the created systems. 

\section{Angular correlations}
A variety of physical phenomena, like the collective behavior of the medium, conservation laws, jets, quantum statistics, or final-state interactions, result in correlations between particles in the final state.

One of the most commonly used experimental techniques are two-particle correlations in relative pseudorapidity ($\Delta\eta$) and azimuthal angle ($\Delta\varphi$) space. The studies typically involve different momentum ranges of particles in the pair: a ``trigger" particle in a certain $p_{\rm T,trig}$ interval and an "associated" particle in a $p_{\rm T,assoc}$ interval. The final correlation is calculated as a per-trigger yield (or ``associated yield per trigger particle"). At RHIC in Au--Au collisions these type of correlations have proved to be a powerful tool to measure and study the properties of high-energy nucleus-nucleus collisions~\cite{Abelev:2009af,Adams:2005ph,Adare:2006nr,Alver:2009id}
Two distinctive features of these measurements were observed: (i) a pronounced peak around $(\Delta\eta,\Delta\varphi)=(0,0)$ originating mostly from jets, called the ``near-side peak", and (ii) a ridge-like correlation structures at $\Delta\varphi=0$ (''near-side") and $\Delta\varphi=\pi$ (``away-side"), elongated over several units of rapidity, usually referred to as the ``ridge". The ridge structure on the near-side, associated with the collective behavior of the medium, has a clear dependence on the centrality of the collision. The same trends have been observed for the Pb--Pb collisions at the LHC~\cite{Aamodt:2011by,Chatrchyan:2012wg}.

\subsection{Double-ridge in p--Pb collisions}
In small systems, like pp or p--A, the shape of the correlation is dominated by the near-side jet peak and the long-range $\Delta\varphi=\pi$ ridge from back-to-back jets. Surprisingly, in high multiplicity pp collisions at $\sqrt{s}=2.76$~TeV, 7~TeV, and 13~TeV at the LHC the similar long-range near-side ridge structure was observed by both CMS and ATLAS detectors~\cite{Khachatryan:2010gv,Aad:2015gqa,Khachatryan:2015lva}. Further analysis of of p--Pb collisions at $\sqrt{s_{\rm NN}}=5.02$~TeV~\cite{CMS:2012qk,Abelev:2012ola,Aad:2014lta,Adam:2015bka} also showed these structures.
Various explanations of these these phenomena have been proposed, either solely based on hydrodynamics (e.g. Refs.~\cite{Bozek:2012gr,Shuryak:2013ke,Bzdak:2013zma}),
or originating from the Color Glass Condensate scenario present in the initial state (e.g. Refs.~\cite{Dusling:2012wy,Altinoluk:2015uaa}).

\begin{figure}[!hbt]
	\centering
	\includegraphics[width=0.49\textwidth]{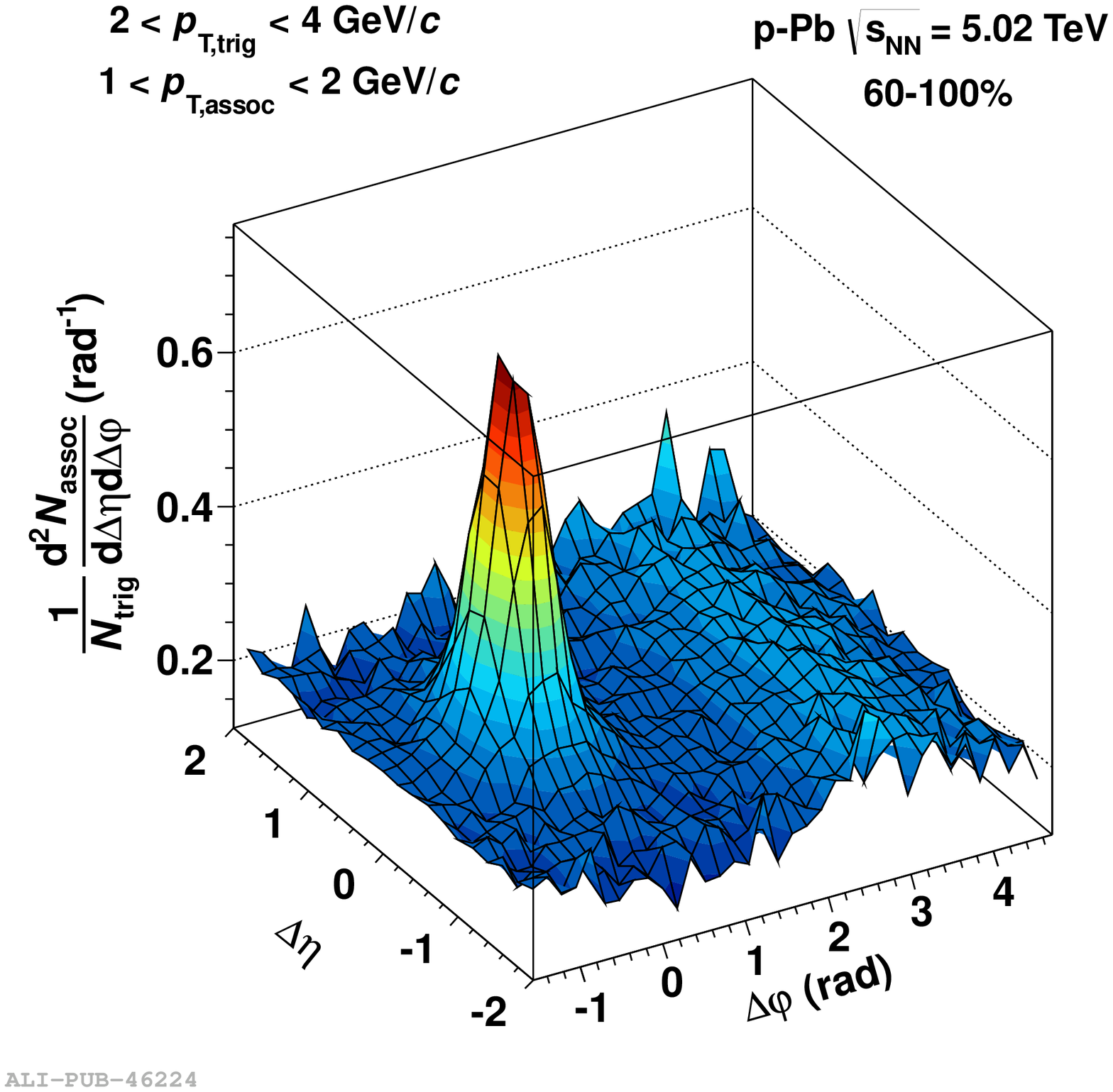}
	\hfill
	\includegraphics[width=0.49\textwidth]{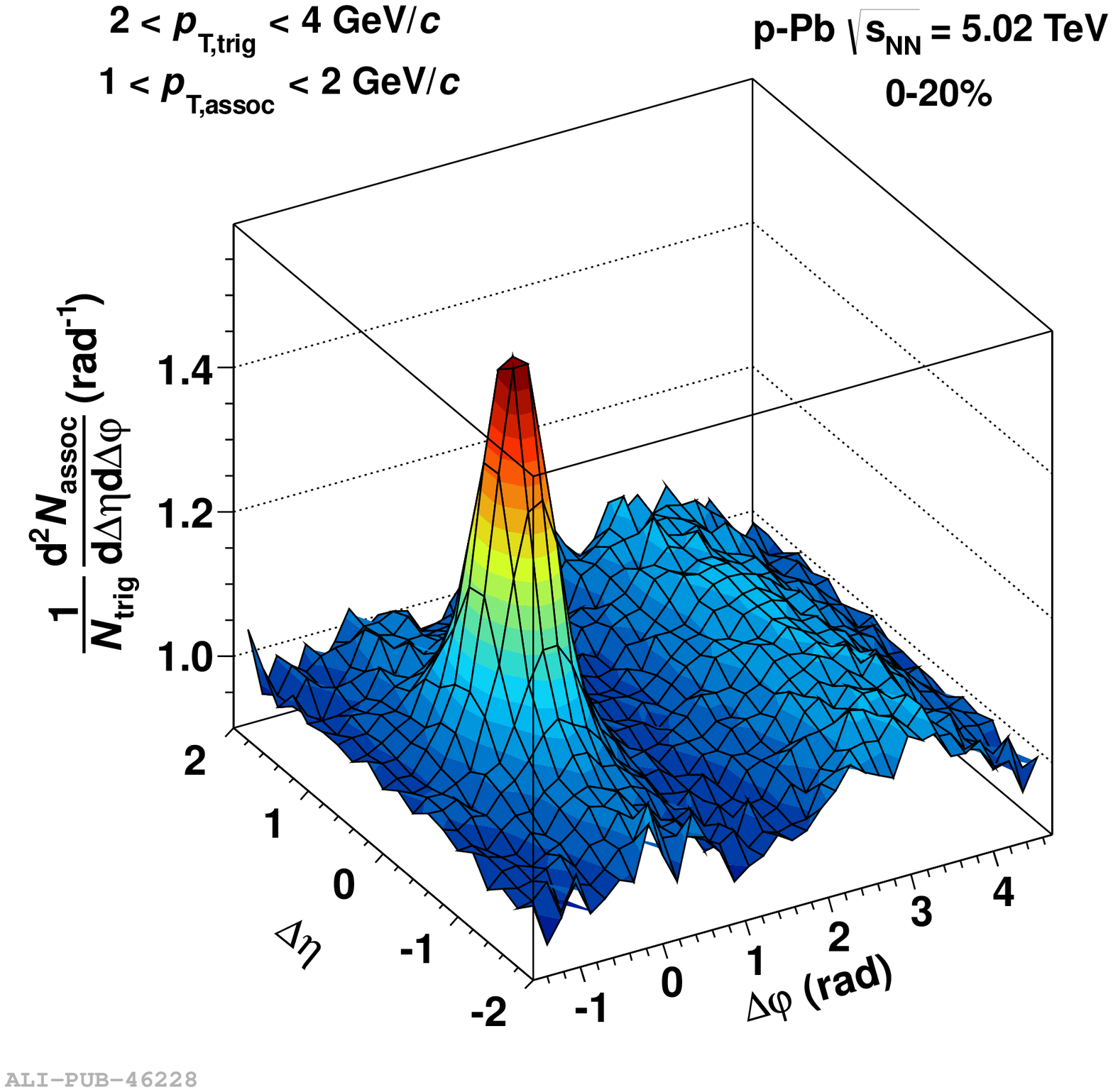}
	\caption{\label{fig:2angcorrs}
		The associated per-trigger yields as a function of $\Delta\varphi$ and $\Delta\eta$ for charged-particle pairs with $2<p_{\rm T,trig}<4$~GeV/$c$ and $1<p_{\rm T,assoc}<2$~GeV/$c$ in p--Pb collisions at $\sqrt{s_{\rm NN}}=5.02$ TeV. Left: results for 60--100\% event class. Right: results for 0--20\% event class~\cite{Abelev:2012ola}.
	}
\end{figure}

\begin{figure}[!hbt]
	\centering
	\includegraphics[width=0.39\textwidth]{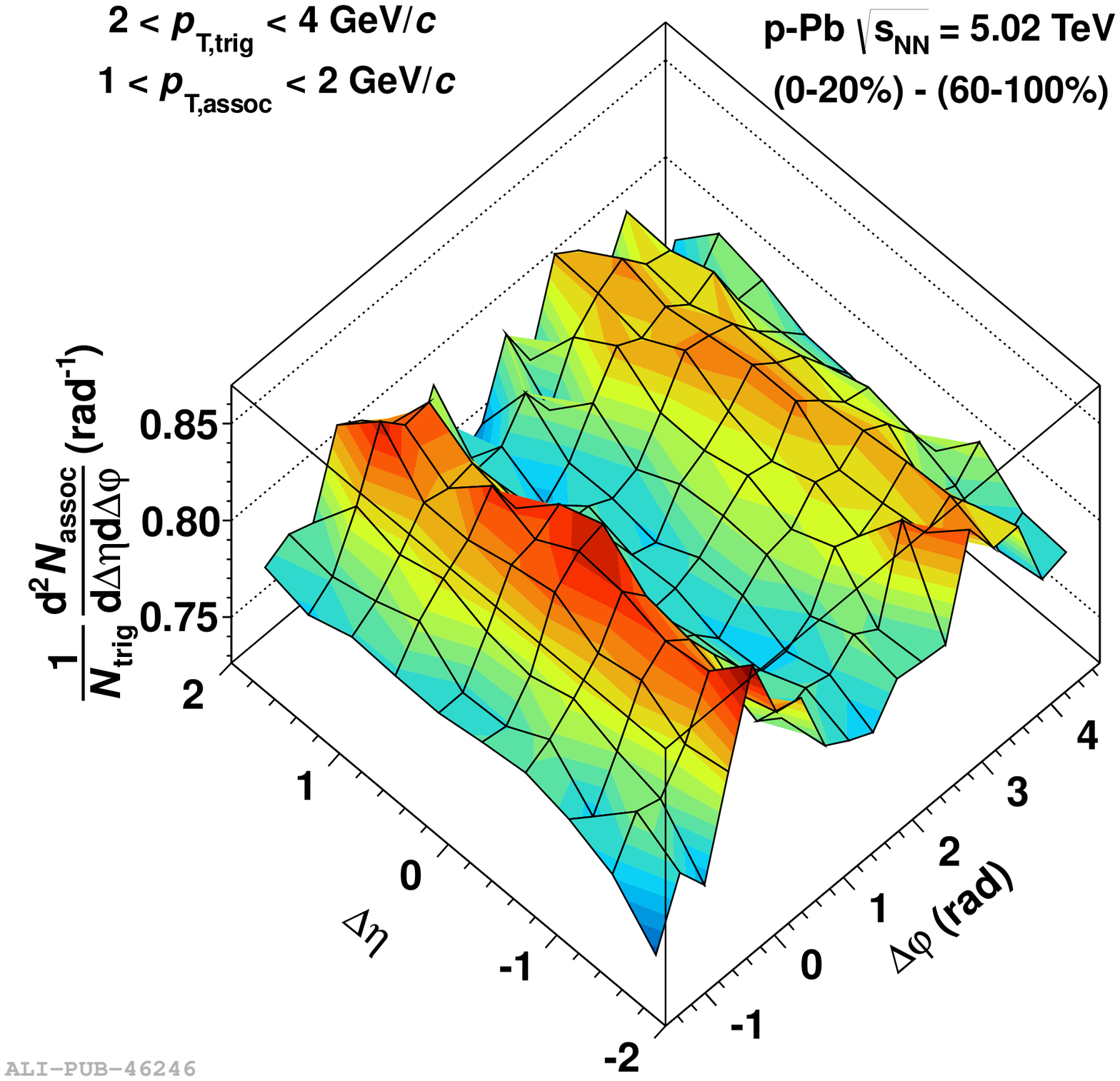}
	\hfill
	\includegraphics[width=0.59\textwidth]{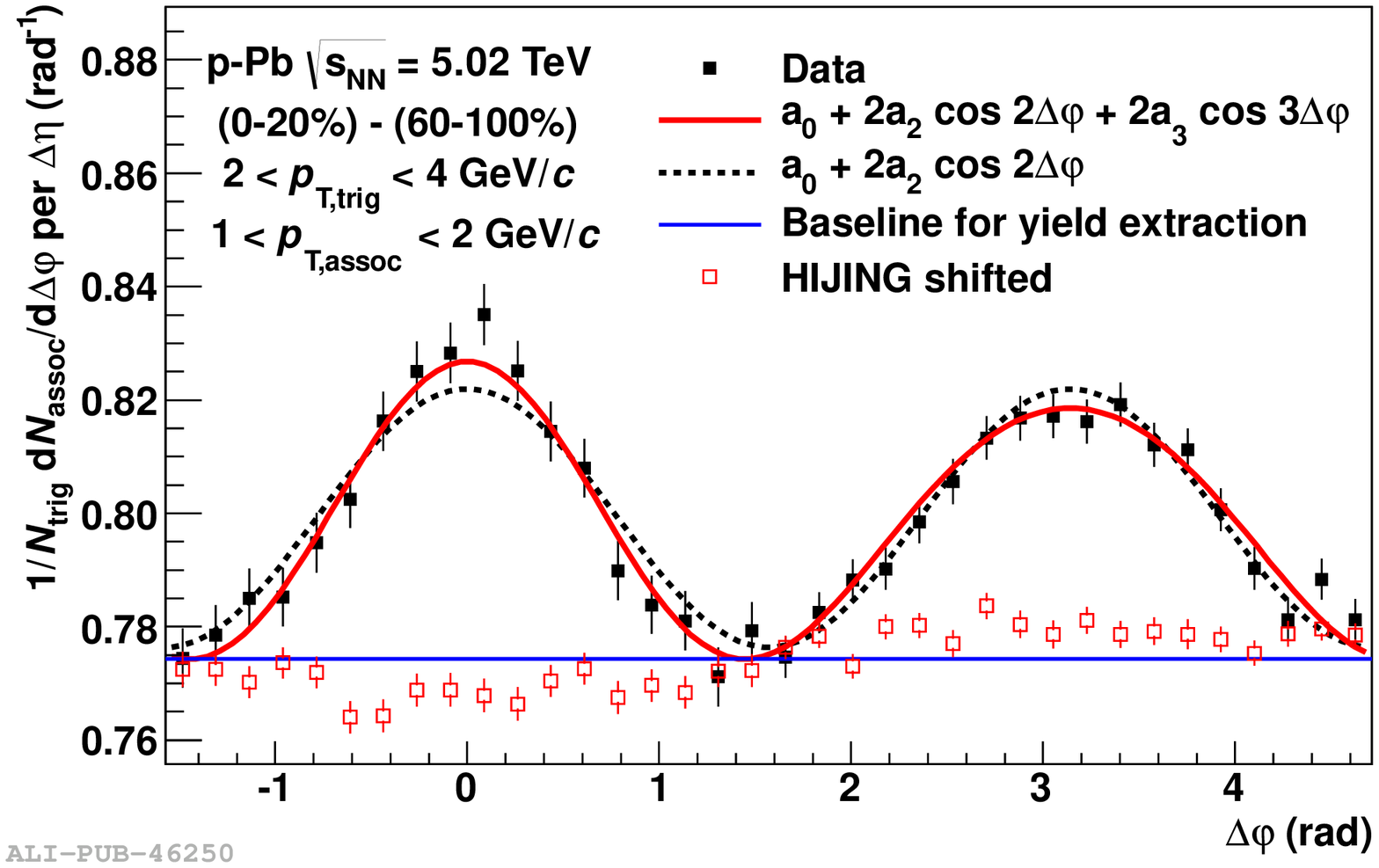}
	\caption{\label{fig:2dsubt}
		Left: the associated per-trigger yield as a function of $\Delta\varphi$ and $\Delta\eta$ for charged-particle pairs with $2<p_{\rm T,trig}<4$ GeV/$c$ and $1<p_{\rm T,assoc}<2$ GeV/$c$ in p--Pb collisions at $\sqrt{s_{\rm NN}}=5.02$~TeV, after subtraction of the associated yield from 60--100\% event class. Right: the associated per-trigger yield after subtraction projected onto $\Delta\varphi$ 
		\cite{Abelev:2012ola}.
	}
\end{figure}

In addition to the near-side ridge structure, the p--Pb results from ALICE revealed another surprising effect -- the presence of two similar long-range ridge-like correlations, one on the near side and one one the away side~\cite{Abelev:2012ola}. This double ridge structure can be observed if the per-trigger yield from the low multiplicity events is subtracted from the high multiplicity events. This procedure approximately removes most of the jet-induced correlations since the near-side yield does not depend on particle multiplicity. The low and high multiplicity per-trigger yields are shown in Fig.~\ref{fig:2angcorrs}. The result of subtraction is shown in Fig.~\ref{fig:2dsubt}-left. One can notice the small peak for $(\Delta\eta,\Delta\varphi)\approx(0,0)$  which still remains. It corresponds to unsubtracted residual jet correlations. Further projections onto $\Delta\varphi$ exclude the region $|\Delta\eta|<0.8$.

The right panel of Fig.~\ref{fig:2dsubt} shows the projection of subtracted per-trigger yield onto $\Delta\varphi$. A modulated signal is clearly observed. We should note that a similar signal extracted from HIJING Monte Carlo model does not show any significant modulation. The modulation effect, for different $p_{\rm T}$ intervals, can be quantified by fitting of the following formula:
\begin{equation}
1/N_{\rm trig} \mathrm{d} N_{\rm assoc}/\mathrm{d}\Delta\varphi = a_0 + 2\,a_2 \cos(2\Delta\varphi) + 2\,a_3 \cos(3\Delta\varphi).
\label{fitfunction}
\end{equation}
The $v_{n}$ flow coefficients can be extracted with the following formula:
\begin{equation}
v_n = \sqrt{a_n / b},
\label{vn}
\end{equation}
where $b$ is the baseline calculated from the high multiplicity event class. This procedure is only possible when $p_{\rm T, trig}$ and $p_{\rm T, assoc}$ intervals are the same. The $v_2$ extracted with this procedure is denoted as $v_2\{\rm 2PC,sub\}$. For the details of the procedure we refer to Ref.~\cite{Abelev:2012ola}.

\begin{figure}[ht!]
	\centering
	\includegraphics[width=0.9\textwidth]{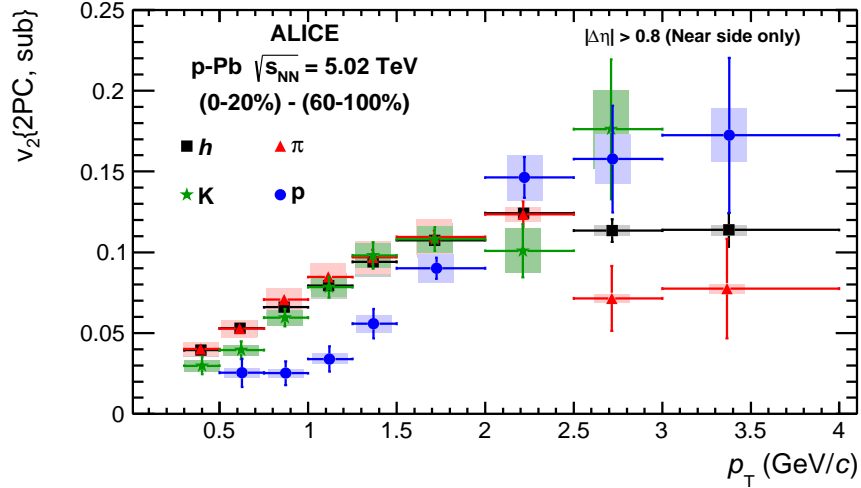}
	\caption{\label{fig:v2_identified}
		The Fourier coefficient $v_2\{\rm 2PC,sub\}$ for hadrons (black squares), pions (red triangles), kaons (green stars), and protons (blue circles) as a function of $p_{\rm T}$ from the two-particle correlation for high-multiplicity collisions after the subtraction of low-multiplicity collisions\cite{ABELEV:2013wsa}.
	}
\end{figure}

A similar subtraction procedure was also applied by ALICE to identified particles (pions, kaons, and protons)~\cite{ABELEV:2013wsa} and the double ridge structure is present as well. The data allowed for the extraction of the flow $v_{2}$ coefficients as a function of $p_{\rm T}$, which are shown in Fig.~\ref{fig:v2_identified}. A clear mass ordering between pions and protons is observed, qualitatively comparable to A--A measurements~\cite{Abelev:2012di}
, which at low $p_{\rm T}$ can be reproduced by models employing hydrodynamic expansion of the medium~\cite{Huovinen:2001cy, Shen:2011eg}. 

Currently, the observation of a long-range double ridge structure in p--Pb collisions is well established by further measurements at the LHC and in d--Au collisions at RHIC (i.e. see Refs.~\cite{Aad:2014lta,Khachatryan:2015waa}).

\subsection{Muon-hadron correlations}
In order to get more insight into the double ridge structure in p--Pb collisions the ALICE collaboration extended the correlation measurements to forward rapidities, taking the advantage of the muon spectrometer located  at pseudorapidity range $-4<\eta<-2.5$~\cite{Adam:2015bka}. In this study, muons were correlated with tracklets\footnote{Tracklets are short track segments reconstructed only with the Silicon Pixel Detector which constitutes the two innermost layers of the Inner Tracking System.} which are measured in the central rapidity region $|\Delta\eta|<1$. In this way particles with $p_{\rm T}$ as low as 50~MeV/$c$ can be detected. The measured muon sample has contribution from decays of pions and kaons, important for $p_{\rm T}<1.5$~GeV/$c$. Above 2 GeV/$c$ muons originate mostly from heavy flavor decays. The p--Pb collisions at $\sqrt{s_{\rm NN}}=5.02$~TeV were delivered by the LHC in two beam configurations: (i) the proton going towards the muon spectrometer (called ``p-going") and (ii) the Pb-ion going in the direction of the muon spectrometer (called ``Pb-going"). Both beam configurations were studied.

\begin{figure}[!hbt]
	\centering
	\includegraphics[width=0.49\textwidth]{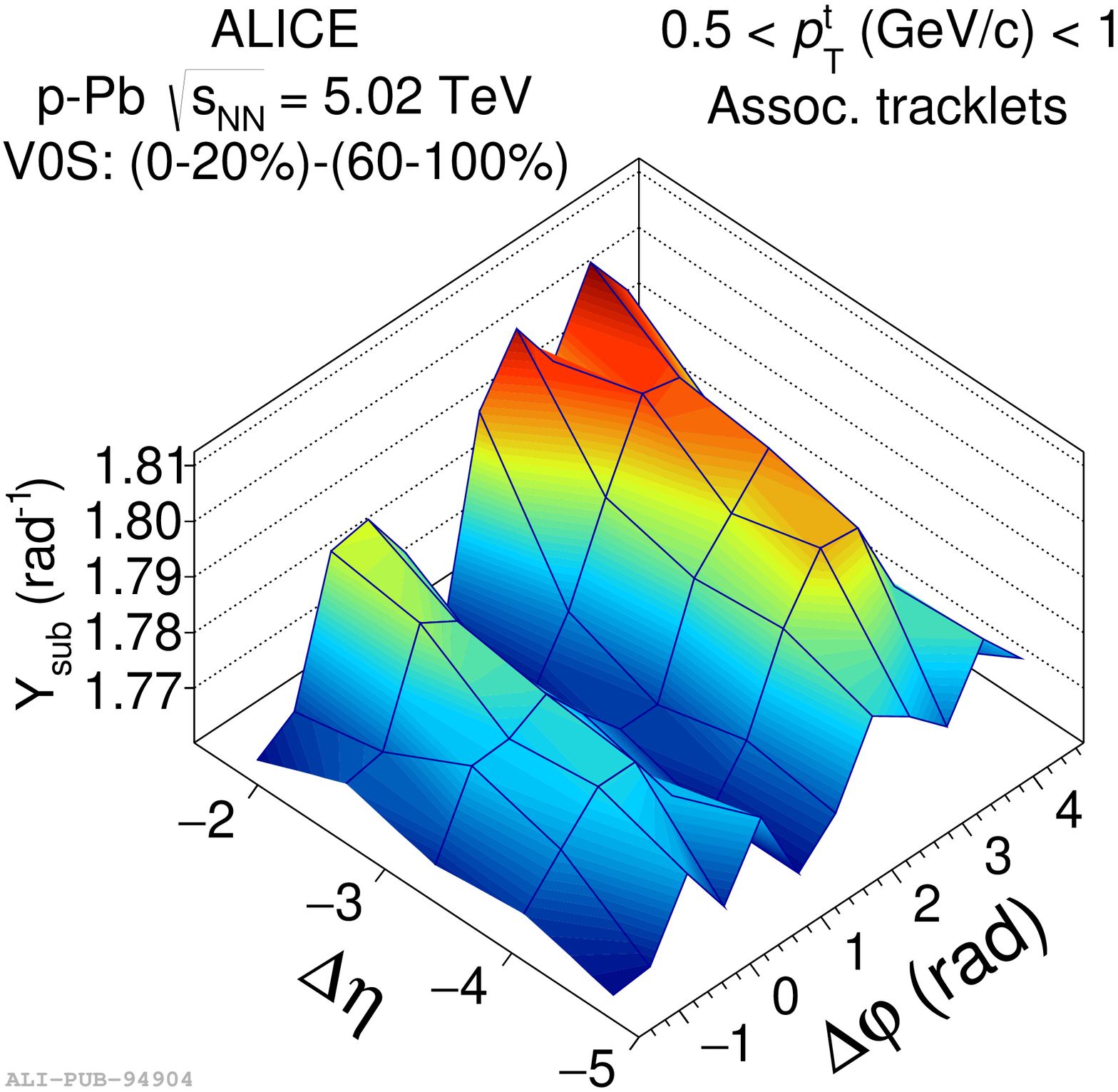}
	\hfill
	\includegraphics[width=0.49\textwidth]{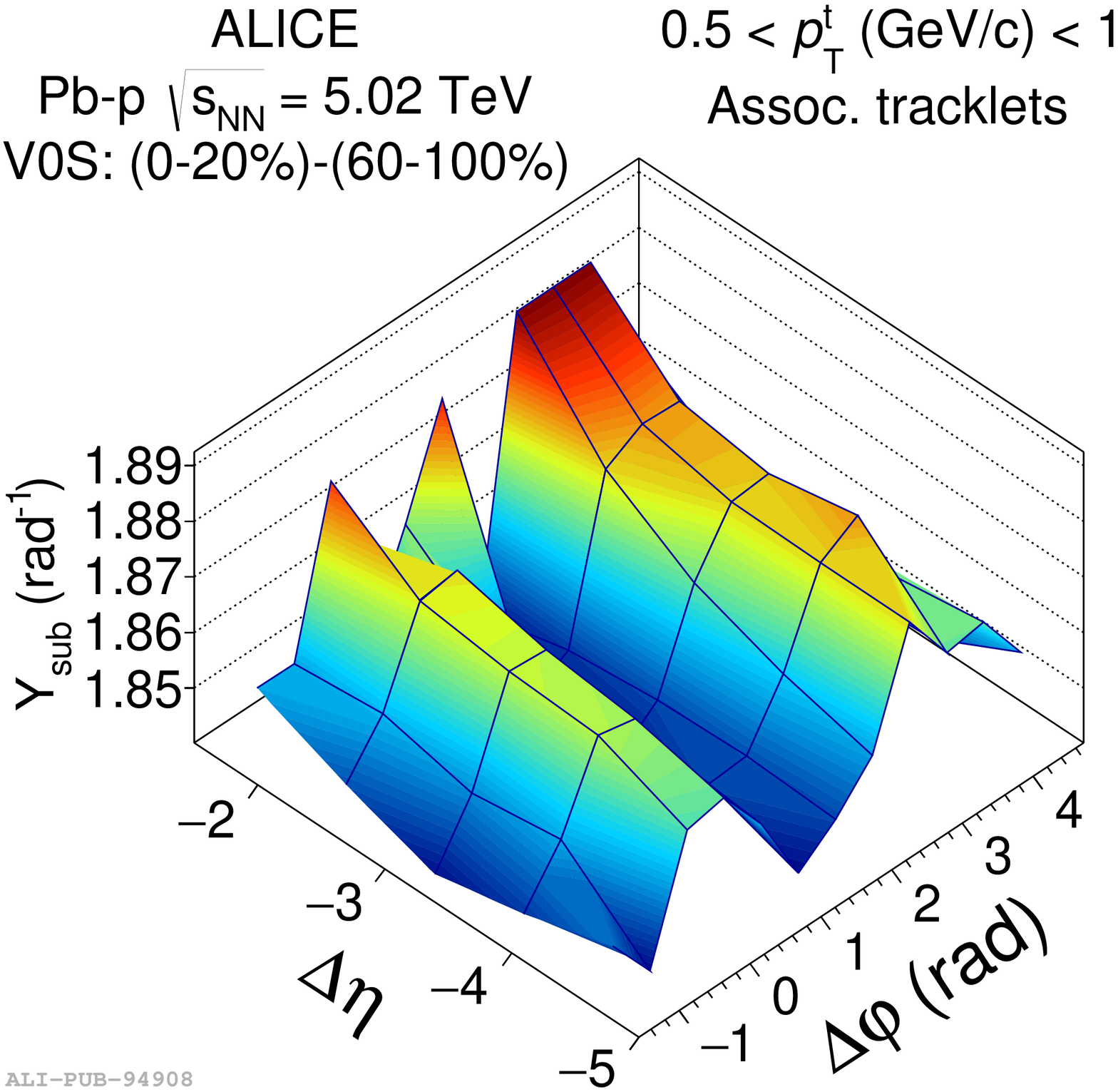}
	\caption{\label{fig:2dsubt_3}
		Muon-hadron correlations in p-going (left panel) and Pb-going (right panel) direction for high-multiplicity collisions after the subtraction of low-multiplicity collisions~\cite{Adam:2015bka}.
	}
\end{figure}

Similarly as in the case of hadron-hadron correlations, the muon-hadron correlations were measured for the highest multiplicity, 0--20\%, and low multiplicity, 60--100\%, event classes. The jet contribution was reduced by subtracting the correlation calculated in 60--100\% event class from the correlation calculated in the 0--20\% event class. The resulting correlation is shown in Fig.~\ref{fig:2dsubt_3}.

\begin{figure}[!hbt]
	\centering
	\includegraphics[width=0.49\textwidth]{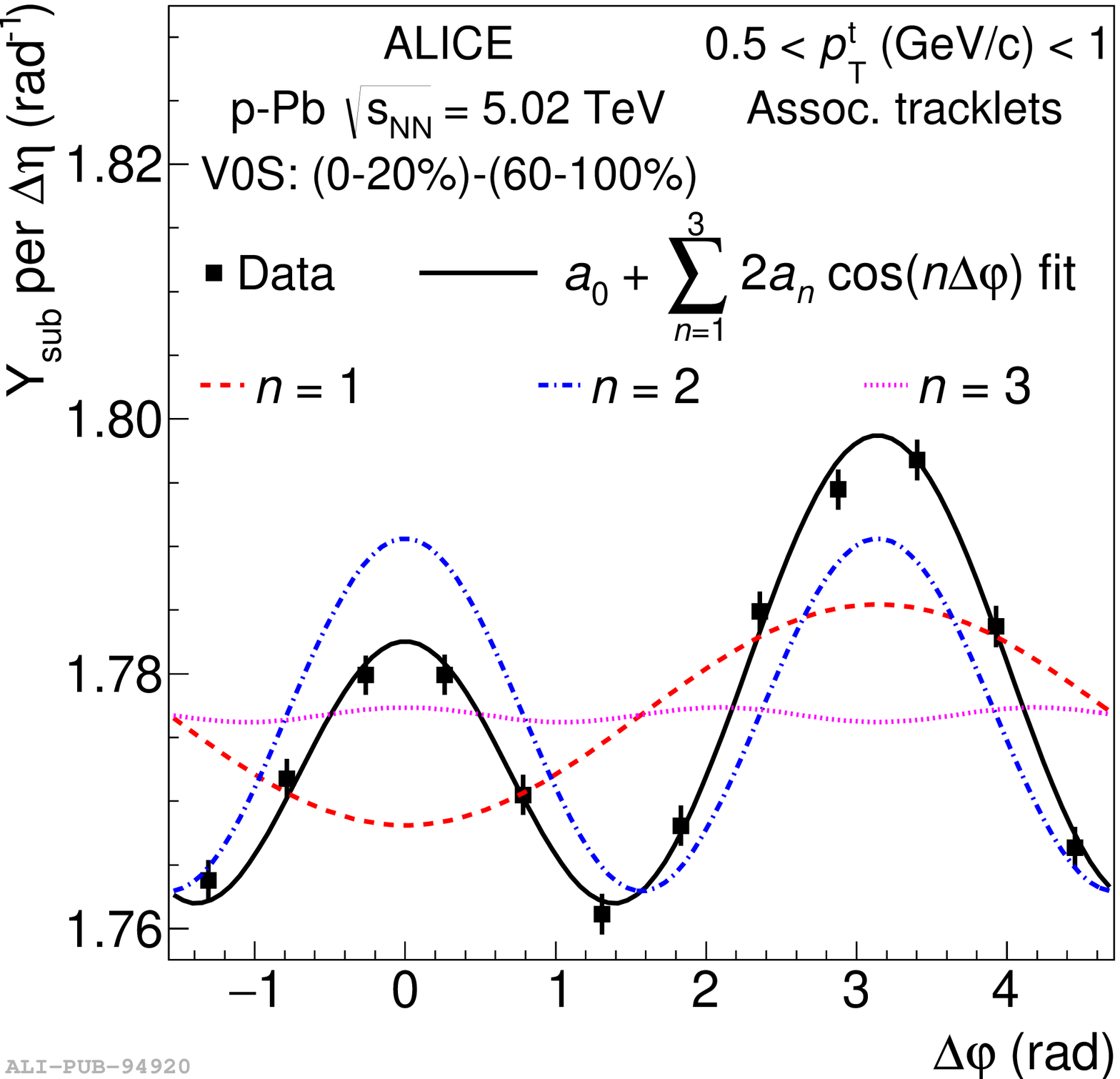}
	\hfill
	\includegraphics[width=0.49\textwidth]{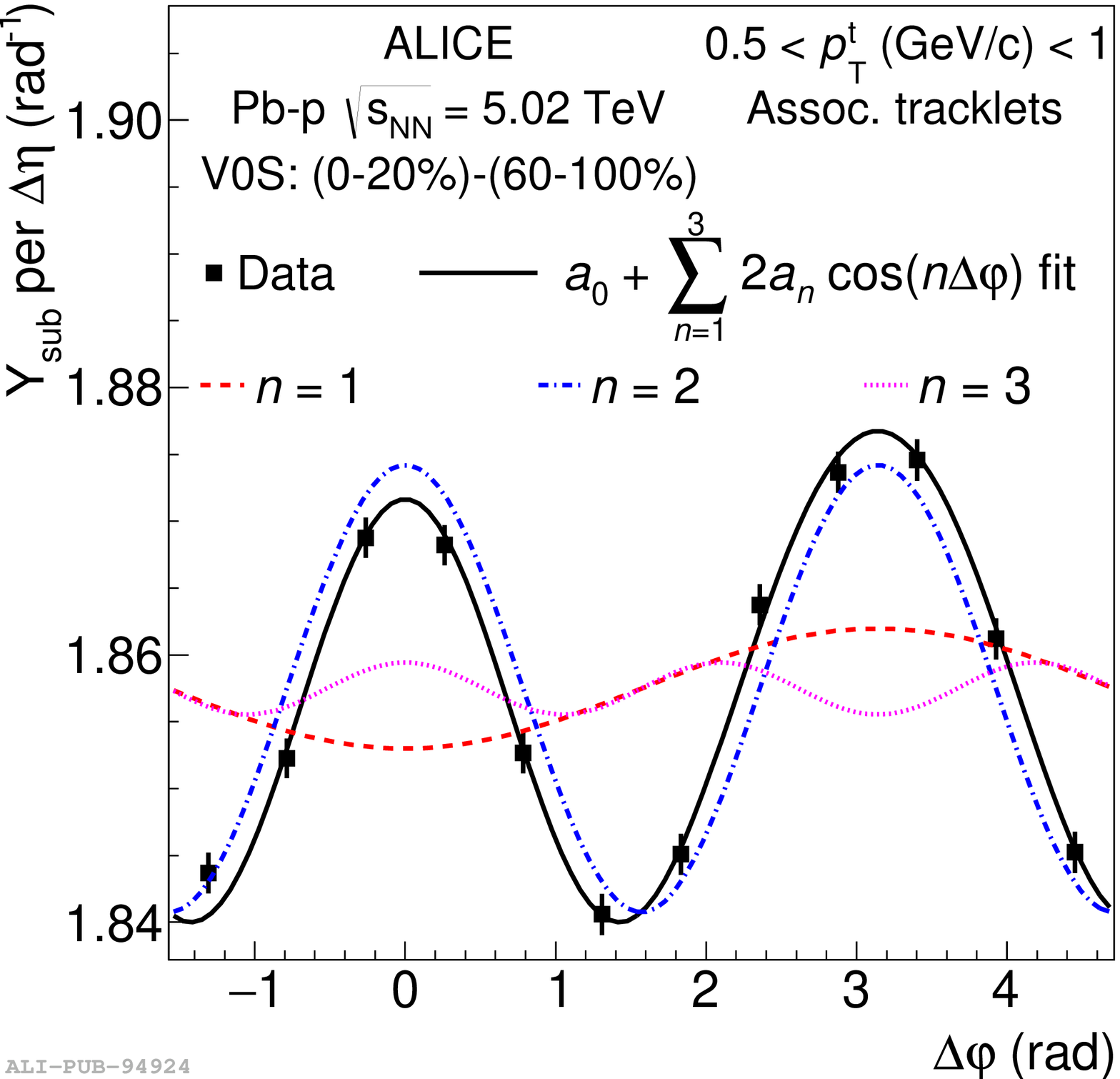}
	\caption{\label{fig:muonProj}
	Projections of muon-hadron correlations to	$\Delta\varphi$ in p-going (left panel) and Pb-going (right panel) direction for high-multiplicity collisions after the subtraction of low-multiplicity collisions. The lines indicate the first three Fourier components of the distribution~\cite{Adam:2015bka}.
	}
\end{figure}

The results show a double ridge structure over 10 units of $\Delta\eta$. Projections on $\Delta\varphi$ of correlations are shown in Fig.~\ref{fig:muonProj}. A Fourier decomposition was applied to the projections:
\begin{equation}
Y_{\rm sub} = a_0 + 2\,a_2 \cos(2\Delta\varphi) + 2\,a_3 \cos(3\Delta\varphi).
\label{fitfunction2}
\end{equation}
The second coefficient dominates in both p-going and Pb-going directions. The result of the decomposition is also shown in Fig.~\ref{fig:muonProj}.

\begin{figure}[!hbt]
	\centering
	\includegraphics[width=0.8\textwidth]{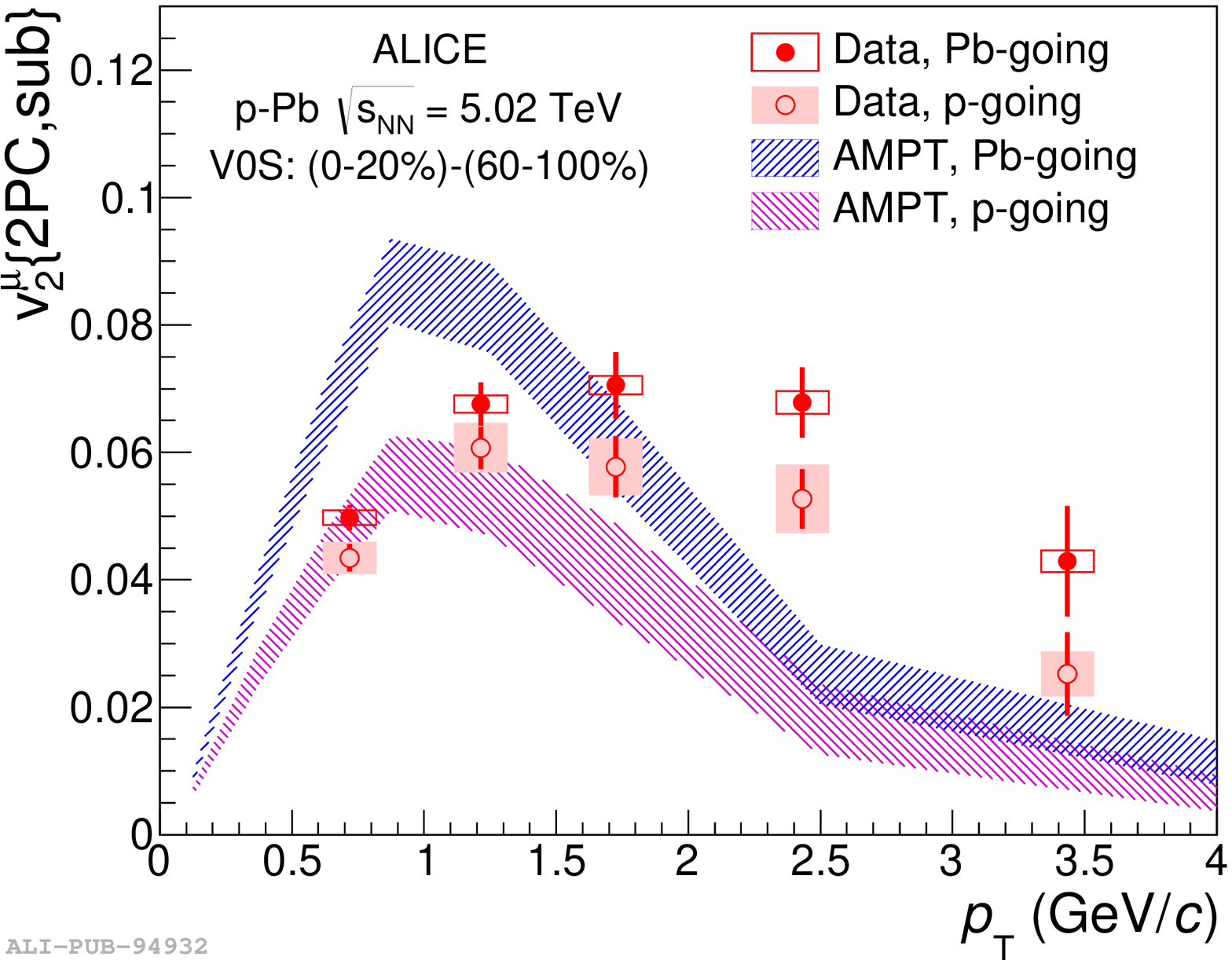}
	\caption{\label{fig:v2_muonProj} 
		$v^{\mu}_{2}\{\rm 2PC,sum\}$ coefficient extracted from muon-hadron correlations after low-multiplicity subtraction (for details see text).  The result from the p-going direction is shown by open symbols, while filled symbols are for Pb-going direction. The result is compared to AMPT predictions~\cite{Adam:2015bka}.
	}
\end{figure}

Figure~\ref{fig:v2_muonProj} presents the extracted $v^{\mu}_{2}\{\rm 2PC,sum\}$ as a function of $p_{\rm T}$. The values are found to be $16\pm6\%$ larger for Pb-going than for p-going direction.
The results from the AMPT model
follow the trend for low $p_{\rm T}$; however, for higher $p_{\rm T}$ they underestimate the data. The higher $v_2$ values at high $p_{\rm T}$, where muons mostly come from heavy-flavor decays, may indicate a non-zero heavy-flavour $v_2$ in the data or a different particle composition in this $p_{\rm T}$ region.

\subsection{Correlations of identified particles in pp collisions}

\begin{figure}[!hbt]
	\centering
	\includegraphics[width=\textwidth]{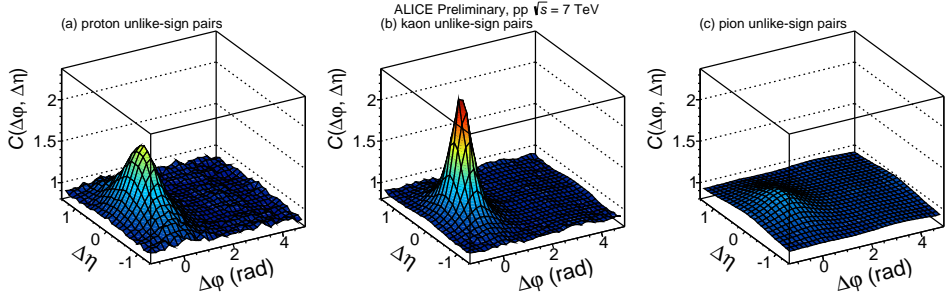}
	\caption{\label{fig:ppUnlike}
	Correlation functions for unlike-sign pairs of protons (left), kaons (middle) and pions (right) for pp at $\sqrt{s}=7$~TeV data.
	}
\end{figure}

\begin{figure}[!hbt]
	\centering
	\includegraphics[width=\textwidth]{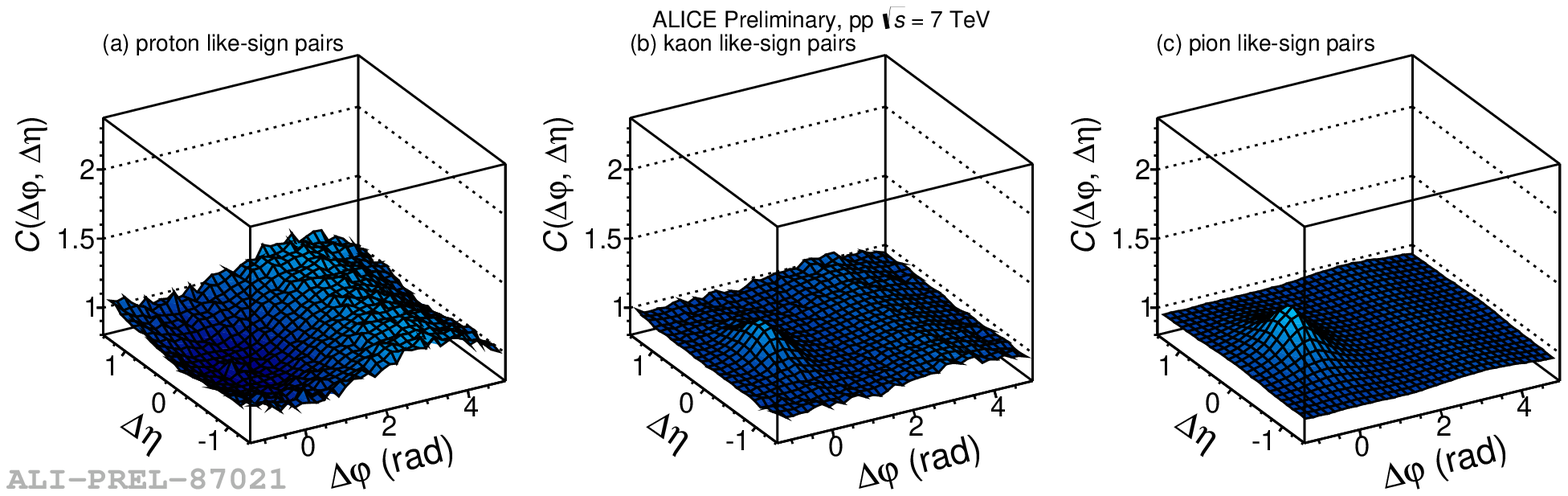}
	\caption{\label{fig:ppLike}
	Correlation functions for like-sign pairs of protons (left), kaons (middle) and pions (right) for pp at $\sqrt{s}=7$~TeV data.
	}
\end{figure}

The correlations analysis in $\Delta\eta$ and $\Delta\varphi$ of identified particles (pions, kaons, and protons) was also performed in pp collisions at $\sqrt{s}=7$~TeV. The measured correlation functions for like- and unlike-sign pairs are presented in Figs.~\ref{fig:ppUnlike} and~\ref{fig:ppLike}. The shape of all correlations modulo like-sign proton pairs show typical near- and away-side structures. While the detailed discussion of the results is available in Ref.~\cite{Graczykowski:2014eqa}, here we would like to focus on the most surprising result -- a wide depression around $(\Delta\eta,\Delta\varphi)=(0,0)$ for like-sign proton pairs. We note that this effect is strictly limited to the baryon-baryon (or antibaryon-antibaryon) scenario. In the case of proton-antiproton correlations the near-side peak is present. 

The following conclusion from this observation can be drawn: baryons are produced in mini-jet fragmentation, however producing more than one baryon-antibaryon pair is strongly suppressed. A similar analysis performed on Monte Carlo data~\cite{Graczykowski:2014eqa} does not reproduce ALICE results. Therefore, the mechanism producing this suppression needs further investigation.

\section{Femtoscopy}

\subsection{Three-dimensional pion femtoscopy}
A technique used to measure the volume of the particle-emitting region at freeze-out is femtoscopy~\cite{Lednicky:2005af,Lisa:2005dd}. In particular, two-pion correlations at low relative momentum $k^{\ast}$ (commonly referred to as Hanbury-Brown, or ``HBT" correlations) have been developed into a precision tool which can be used to extract a detailed information system size and its dependence on event multiplicity and pair transverse momentum, $k_{\rm T}$. Femtoscopy, in general, measures the width of the distribution of relative separation between the emission points of two particles, which is conventionally referred to as the ``radius parameter" (or the ``HBT radius"), and can be evaluated in three dimensions: \emph{long} along the beam axis, \emph{out} along the pair transverse momentum, and \emph{side} perpendicular to the other two. Moreover, the femtoscopic results are usually interpreted within the hydrodynamic framework as a signature of collective behavior of the strongly interacting medium and provide crucial constraints on phase transition to hadronic matter. ALICE results on three-dimensional pion femtoscopy in pp, p--Pb and Pb--Pb collisions can be found in~\cite{Aamodt:2011mr,Aamodt:2010jj,Aamodt:2011kd,Adam:2015pya,Adam:2015vna}.

\begin{figure}[!hbt]
	\centering
	\includegraphics[width=0.49\textwidth]{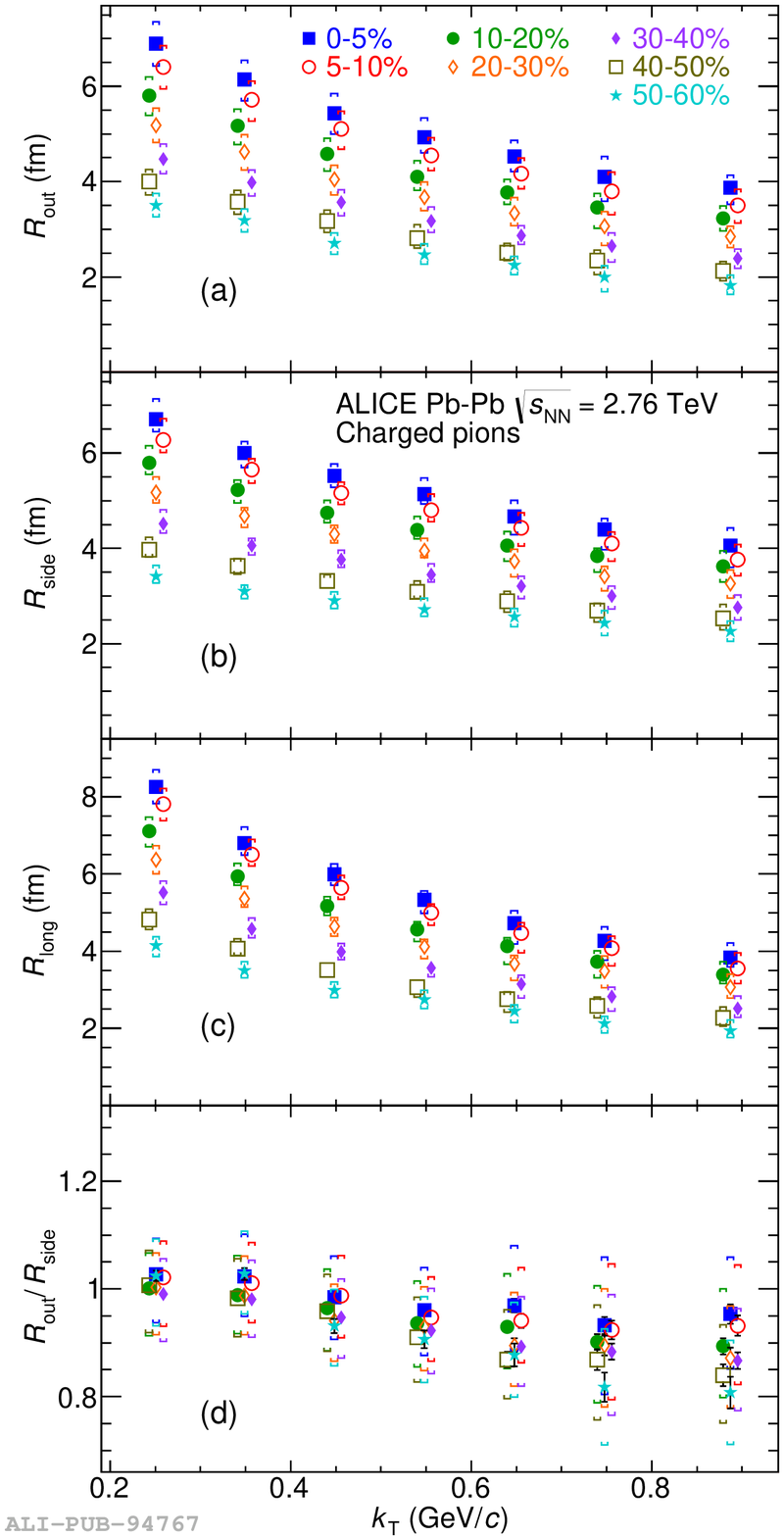}
	\hfill
	\includegraphics[width=0.49\textwidth]{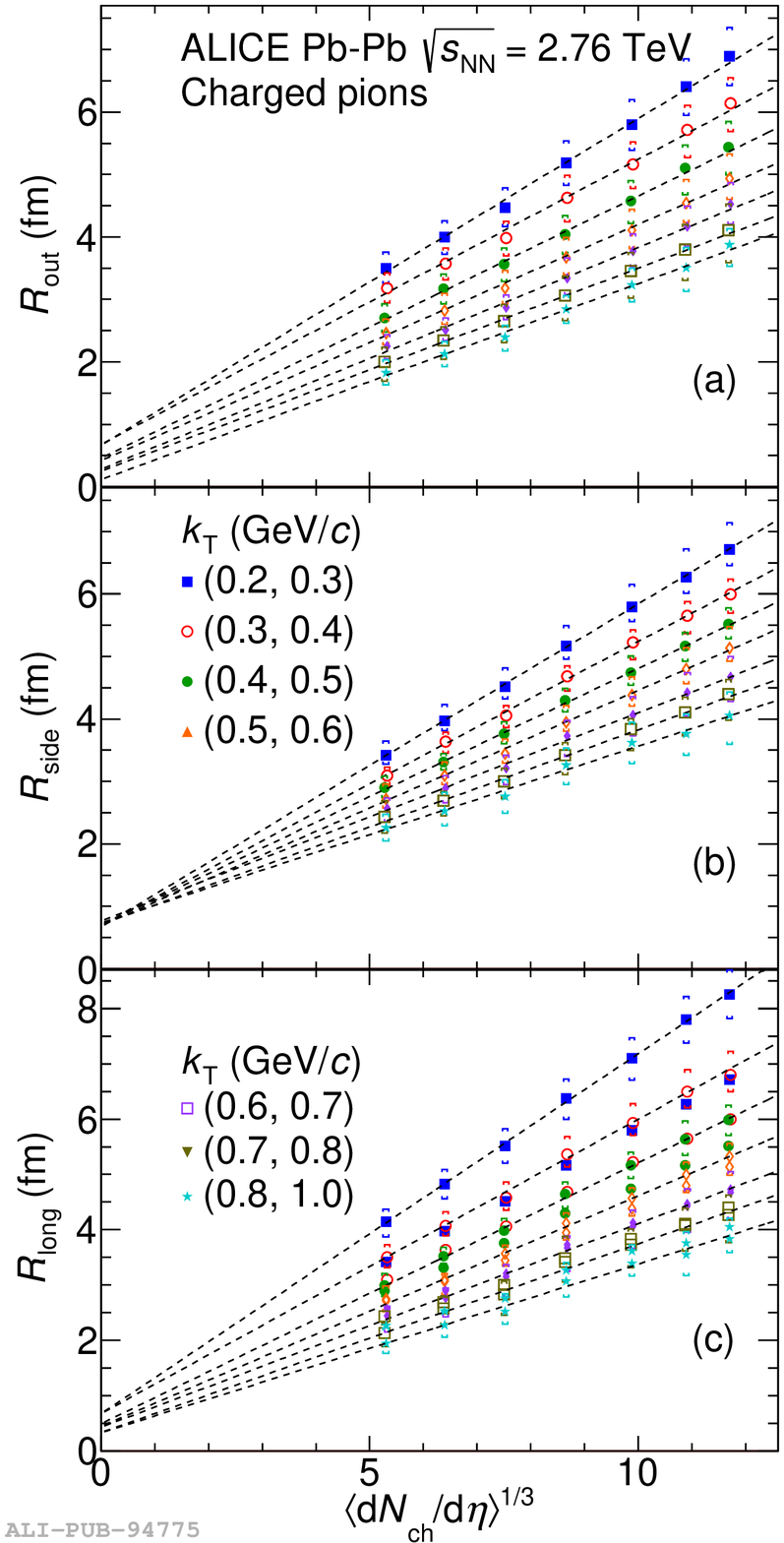}
	\caption{\label{fig:centralityFemtoPbPb}
		Left: femtoscopic radii for seven centrality ranges shown as a function of pair transverse momentum $k_{\rm T}$. Right: femtoscopic radii shown as a function of the cube root of charged particle multiplicity density. For better visibility some points were shifted in $x$ direction~\cite{Adam:2015vna}.
	}
\end{figure}

In heavy-ion collisions two clear trends can be observed: (i) all three radii scale approximately linearly with the cube root of the charged particle multiplicity density at midrapidity, $\langle {\rm d}N_{\rm ch}/{\rm d}\eta \rangle^{1/3}$, and (ii) they decrease with pair transverse momentum. The ALICE results from Pb--Pb data at $\sqrt{s_{\rm NN}}=2.76$~TeV for different centrality and $k_{\rm T}$ ranges~\cite{Adam:2015vna}, showing both trends, are presented in Fig.~\ref{fig:centralityFemtoPbPb}.

\begin{figure}[!hbt]
	\centering
	\includegraphics[width=\textwidth]{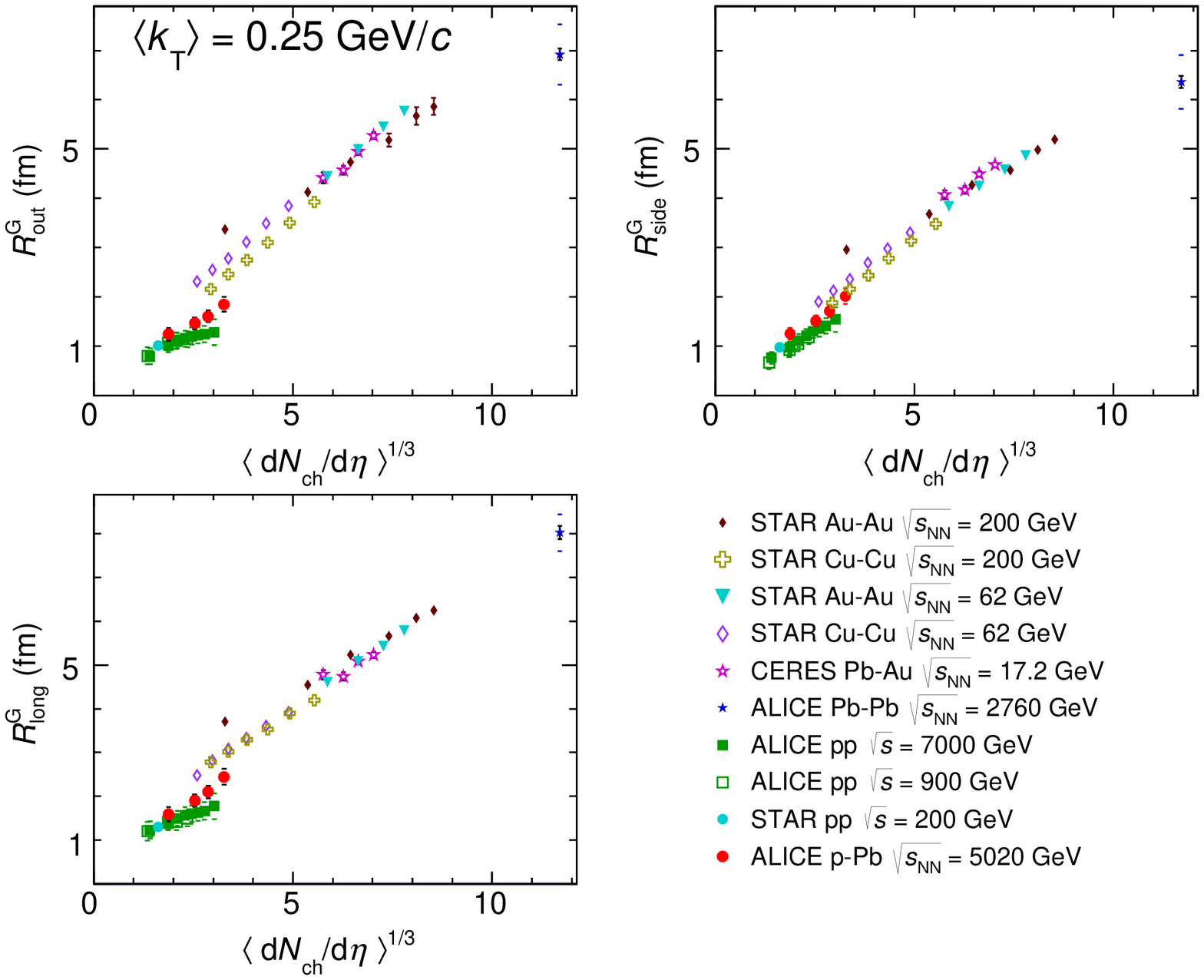}
	\caption{\label{fig:WorldDataFemto}
	Femtoscopic radii from various collision systems and energies as a function of cube root of the measured charged-particle multiplicity density~\cite{Adam:2015pya}.
	}
\end{figure}

It has been argued in Ref.~\cite{Lisa:2005dd}, that not only the three-dimensional radii scale with the cube root of the multiplicity density for a single collision energy, but across different energies and colliding system. Indeed, one can clearly see (Fig.~\ref{fig:WorldDataFemto}) significantly different scaling between A--A and pp systems. The p--Pb radii tend to agree with those in pp at low mulitplicities and start to diverge for increasing multiplicities. This finding is confirmed with the three-pion cumulant correlation analysis performed in all three systems~\cite{Abelev:2014pja}.

\subsection{$\mathrm{K}^0_{\rm S}\mathrm{K}^{\pm}$ femtoscopy}
The femtoscopic formalism is not limited to pions only. Recently, results of identical-kaon (neutral and charge) femtoscopy have been published by the STAR Collaboration for Au--Au collisions at $\sqrt{s_{\rm NN}}=0.2$~TeV~\cite{Abelev:2006gu} as well as for pp data at $\sqrt{s}=7$~TeV and Pb--Pb collisions at $\sqrt{s_{\rm NN}}=2.76$~TeV by the ALICE Collaboration~\cite{Abelev:2012ms,Abelev:2012sq,Adam:2015vja}. The correlation function is a result of the interplay of the following phenomena: quantum statistics (for $\mathrm{K}^{\pm}\mathrm{K}^{\pm}$ and $\mathrm{K}^0_{\rm S}\mathrm{K}^0_{\rm S}$), Coulomb interaction ($\mathrm{K}^{\pm}\mathrm{K}^{\pm}$), and the final-state interaction through the $f_0(980)/a_0(980)$ threshold resonances (for $\mathrm{K}^0_{\rm S}\mathrm{K}^0_{\rm S}$). In addition to identical-kaon system, $\mathrm{K}^0_{\rm S}\mathrm{K}^{\pm}$ correlations can also be considered, though no such measurements have been performed before this study. For these correlations, in addition to the trivial elastic scattering channel, the only allowed final-state pair-wise interaction proceeds through the $a_0(980)$ resonance\footnote{The $\mathrm{K}^0_{\rm S}\mathrm{K}^0_{\rm S}$ pair is in $I=1$ isospin state, as is the $a_0$, whereas the $f_0$ is in $I=0$ state, so the isospin would not be conserved.}.

Another property of the $\mathrm{K}^0_{\rm S}\mathrm{K}^{\pm}$ interaction through the $a_0$ resonance is also the fact that the $a_0$ has strangeness $S=0$. The $\mathrm{K}^0_{\rm S}$ state is a linear combination of $\mathrm{K}^0$ and $\mathrm{\bar{K^0}}$ states. In order to conserve strangeness only the $\mathrm{\bar{K^0}K^+}$ pair from $\mathrm{K^0_SK^+}$ and the $\mathrm{\bar{K^0}K^-}$ pair from $\mathrm{K^0_SK^-}$ can form the $a_0$. This feature allows the possibility to study the $\rm K^0$ and $\rm \bar{K^0}$ sources separately.

In addition to above possibilities the $\mathrm{K}^0_{\rm S}\mathrm{K}^{\pm}$ final-state interaction allows the study of the properties of the $a_0$ itself. Its interest comes from the fact that many papers in the literature discuss the possible scenario that the $a_0$ resonance could be a 4-quark state, i.e. a tetraquark, or a ``$\rm \bar{K}-K$ molecule"~\cite{Martin:1976vx,Antonelli:2002ip,Achasov:2001cj,Achasov:2002ir}. 

\begin{figure}[!hbt]
	\centering
	\includegraphics[width=\textwidth]{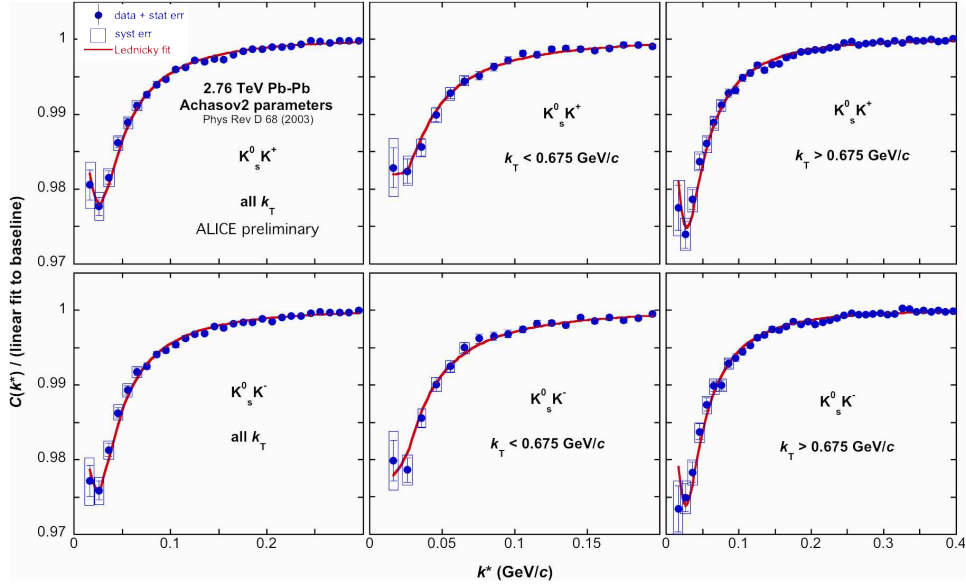}
	\caption{\label{fig:K0sK+-}
	Examples of $\mathrm{K^0_S}\mathrm{K^+}$ and $\mathrm{K^0_S}\mathrm{K^-}$ correlation functions and fit with the Lednicky parametrization using "Achasov2"~\cite{Achasov:2001cj} parameters.
	}
\end{figure}

Figure~\ref{fig:K0sK+-} shows examples of $\mathrm{K^0_S}\mathrm{K^+}$ and $\mathrm{K^0_S}\mathrm{K^-}$ correlation functions with Lednicky fits~\cite{Lednicky:1981su,Bekele:2007zza} using the ``Achasov2"~\cite{Achasov:2001cj} parameters of the $a_0$. The main feature of the femtoscopic correlation function can be observed: the suppression caused by the strong final-state interactions for small $k^{\ast}$. From this plot we can conclude that the $a_0$ final-state interaction gives an excellent representation of the the data, i.e. the suppression of the correlation functions in the $k^{\ast}$ range up to $0.15$~GeV/$c$.

\begin{figure}[!hbt]
	\centering
	\includegraphics[width=\textwidth]{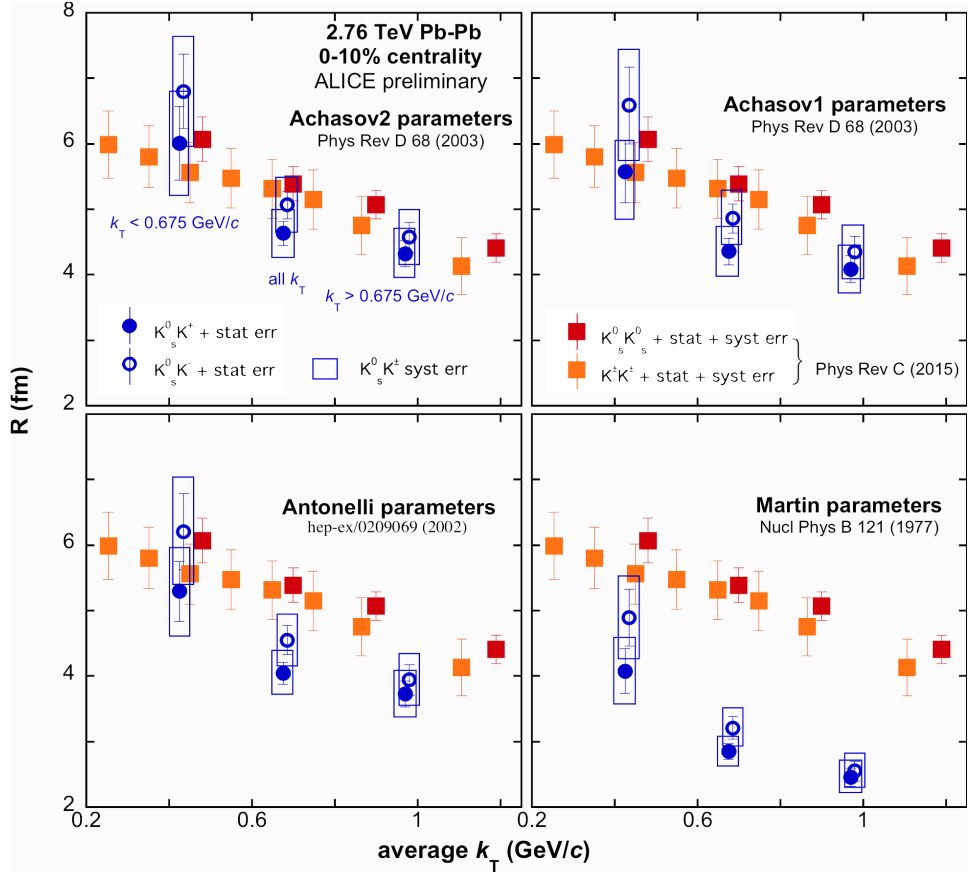}
	\caption{\label{fig:RK0sK+-}
	$R$ fit parameters from averaged $\mathrm{K}^{0}_{\rm S}\mathrm{K}^{\pm}$ analysis compared to identical kaon femtoscopy from ALICE~\cite{Adam:2015vja}.
	}
\end{figure}

\begin{figure}[!hbt]
	\centering
	\includegraphics[width=\textwidth]{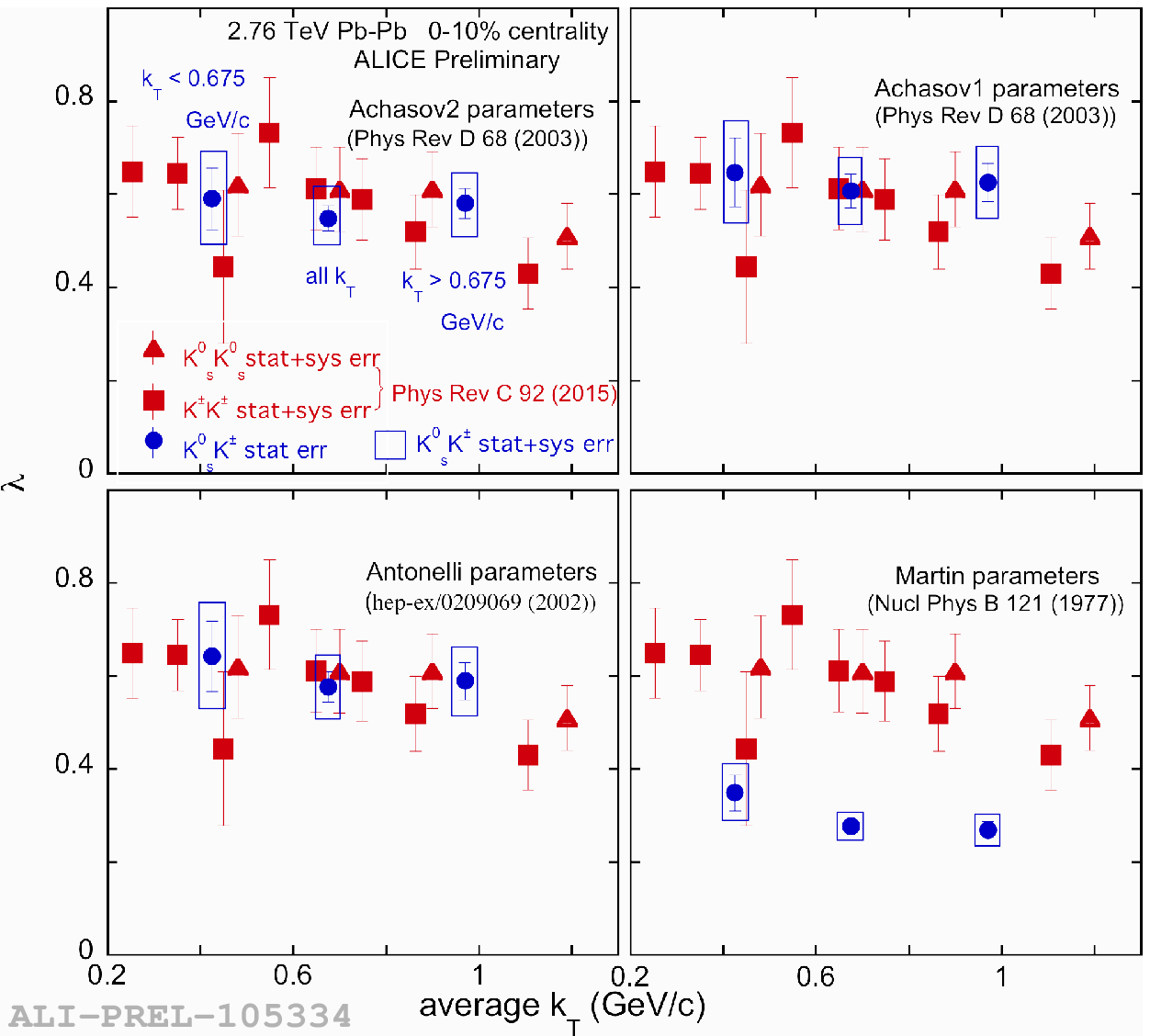}
	\caption{\label{fig:LamK0sK+-}
	$\lambda$ fit parameters from averaged $\mathrm{K}^{0}_{\rm S}\mathrm{K}^{\pm}$ analysis compared to identical kaon femtoscopy from ALICE~\cite{Adam:2015vja}.
	}
\end{figure}

The results of the fit, $R$ (the size of the kaon source) and $\lambda$ (the strength of the correlation) parameters, for all considered $a_0$ parameterizations (in the decreasing order from the largest to the lowest $a_0$ parameters: ``Achasov2"~\cite{Achasov:2002ir}, ``Achasov1"~\cite{Achasov:2001cj}, ``Antonelli"~\cite{Antonelli:2002ip}, and ``Martin"\cite{Martin:1976vx}) are presented in Figs.~\ref{fig:RK0sK+-} and~\ref{fig:LamK0sK+-}. Since  $\mathrm{K^0_S}\mathrm{K}^{+}$ and $\mathrm{K^0_S}\mathrm{K}^{-}$ are consistent with each other, both parameters shown are calculated as their average.

The comparison of the radius parameter with identical kaon results shows clear agreement with each other for ``Aachasov", ``Achasov2", and ``Antonelli" parameterizations of $a_0$ resonance. This is expected from the fact that radii from both $\mathrm{K}^{0}_{\rm S}\mathrm{K}^{0}_{\rm S}$ and $\mathrm{K}^{\pm}\mathrm{K}^{\pm}$ pair combinations have similar source geometry and there is no reason for $\mathrm{K}^{0}_{\rm S}\mathrm{K}^{\pm}$ to be different. A clear discrepancy is visible for ``Martin", which corresponds to the lower values of $a_0$ parameters. Therefore, the higher values of $a_0$ parameters are favored.

The $\lambda$ parameters of identical kaon results also agree with $\mathrm{K}^{0}_{\rm S}\mathrm{K}^{\pm}$. This is consistent with the assumption of 100\% final-state interaction going through the $a_0$ resonance channel.

\section{Conclusions}
Several recent correlation results from ALICE have been presented. The angular correlations analysis in p--Pb collisions revealed the existence of a double ridge structure, similar to the one observed in A--A data and usually interpreted as a signature of collectivity. The measurements were extended with correlations at forward rapidities thanks to ALICE muon spectrometer. Similar correlation studies in pp reveal surprising anti-correlation structure for identical proton pairs, which is not reproduced by existing models. This result suggests strong influence of local conservation laws on the shape of the observed correlation.

The pion femtoscopic analysis in Pb--Pb show clear multiplicity and pair transverse momentum scalings for all three radii. Comparison of A--A, p--A, and pp radii as a function of cube root of charged particle multiplicity density across various experiments and collision energies show a universal trend for A--A data, different from the one observed in pp collisions. The p--Pb results from ALICE tend to agree with those of pp at low multiplicites and start to diverge as the multiplicity increases.
\\

The author wishes to acknowledge the financial support of the Polish National Science Centre under decisions no. 2013/08/M/ST2/00598 and no. UMO-2014/13/B/ST2/04054.

\bibliographystyle{h-physrev}
\bibliography{bibliography}


\end{document}